**First-principles Studies on Structural, Electronic, Optical and Mechanical Properties of Inorganic $CS_2NaTlX_6$ (X = F, Cl, Br) Double Halide Perovskites.**

Mohammed Mehedi Hasan[a], Nazmul Hasan[b], Alamgir Kabir[c]
[a]Department of Theoretical Physics, University of Dhaka, Dhaka-1000, Bangladesh
[b]Department of Electrical and Computer Engineering, North South University, Dhaka-1229, Bangladesh
[c]Department of Physics, University of Dhaka, Dhaka-1000, Bangladesh
a mehedi.tpdu@gmail.com
b nzharis97@gmail.com
c alamgir.kabir@du.ac.bd

Abstract:

Lead-free double halide perovskites have shown various benefits over the lead-based perovskite, due to their suitable optical absorption efficiency, higher stability, tunable bandgap, large carrier mobility, non-toxicity, easily available raw materials, low cost, etc The structural, electrical, optical, and mechanical characteristics of the lead-free halide double perovskites $Cs_2NaTlX_6$ (X = F, Cl, Br) are calculated by utilizing Perdew–Burke–Ernzerhof (PBE) functional within generalized gradient approximation (GGA) under the context of density functional theory (DFT). The structural properties such as lattice parameter, cell volume, total energy, bulk modulus, pressure derivative, and tolerance factor are computed at equilibrium. The electronic density of states reveal the semiconducting nature of the compound and the band structure exhibit the nature of the band gap to be direct. HSE06 functional is introduced to correct the under-estimated band gap as obtained in the GGA-PBE functional. The real and imaginary components of the dielectric function, absorption coefficient, energy loss function, reflectivity, refractive index, and extinction coefficient are analyzed and explained by electronic structures. These parameters indicate that Br and Cl-based materials are optically more suitable than F-based compound. The mechanical properties ensure the ductile nature of the halide double perovskites.

**Introduction:**

Perovskites are excellent candidates for sustainable and renewable energy applications, such as photovoltaics, due to their growing power conversion efficiency (PCE) from 3.8% as reported in 2009 [1] to 23.7% in 2020 [2], [3] when they are used as active material in solar cells. The remarkable intrinsic characteristics of perovskite solar cells, including their large amount of visible-light absorption [4], [5], the small binding energy of exciton [6], [7], long carrier diffusion length [8], [9], long carrier lifetime and high carrier mobility [10], [11], are the reason for their large value of PCE. These materials are very important for the scientific community to meet the increased global energy demands over traditional energy sources and climate change challenges [12]. However, the low stability and the toxicity of the Pb-based materials are the two main challenges to commercializing lead-based perovskite for photovoltaic applications [13]. Therefore, lead-free double perovskites (DPs), a similar geometrically shaped compound derived from the ideal perovskite structure $ABX_3$, are the potential candidate to address these issues because of their high stability [14], [15] and environmentally friendly role against lead toxicity [16], [17]. A DP has a general formula of $A_2BB'X_6$, where A represents a large element (Cs or Rb), B and B' are monovalent and trivalent cations respectively (B = $Na^+$, $Ag^+$, $Cu^+$; B' = $Bi^{3+}$, $Sb^{3+}$, $Tl^{3+}$, $In^{3+}$), and X represents a halide ion (X = $F^-$, $Cl^-$, $Br^-$) [17], [18]. These materials have low carrier effective masses and highly tunable band gaps in the range of visible light spectrum.

Halide DPs show both indirect and direct band gaps and several studies of different combinations of DPs such as $Cs_2AgBiBr_6$, $Cs_2AgBiCl_6$, $Cs_2NaSbCl_6$, and $Cs_2CuSbCl_6$ have reported indirect band gaps [19]–[22]. On the other hand, direct bandgap semiconductors, which are used in the photovoltaic applications, because it has small energy loss, be more readily inspired, and be better capable of completing the photoelectric conversion [23]. Over the years, many halide DPs have been investigated both theoretically and experimentally including $Cs_2AgInCl_6$, $Cs_2NaInCl_6$, $Cs_2AgTlX_6$ (X = Cl, Br), Cs2InSbCl6 as well as the alloyed structures of these compound exhibit direct band gap and excellent PCE [24]–[30]. Recently, a new theoretical study has reported the stability of $Cs_2NaTlBr_6$ double perovskite, they have found a direct bandgap of 1.82 eV which can be a suitable candidate for photovoltaic solar cell [31].

To date, there has been a lack of comprehensive and comparative studies that have investigated the effect of halogen substitution on the properties of NaTl-containing double halide perovskites. To address this gap, in this study we aimed to investigate the structural, electronic, optical, and mechanical properties of halide double perovskites Cs2NaTlX6 (X= F, Cl, Br) through first-principle calculations using the GGA-PBE method (exchange-correlation energy of a system using the gradient of the electron density), which is based on density functional theory (DFT). Furthermore, we employed the hybrid functional HSE06 method to obtain more accurate estimates of the band gap values surpassing the limitations of GGA-PBE functional that underestimates the band gaps of semiconductors and insulators. Our results showed that the band gap values of the compounds decreased as the atomic size of the halogen element increased. The structural and mechanical properties of the material were found to be stable. Furthermore, we compared our results with those of previous theoretical and experimental studies and found a good match. The results of our study will provide new insights into the potential of lead-free halide DP materials for optoelectronic applications of NaTl-based double halide perovskites.

**Computational methodology:**

The Vienna ab initio simulation package (VASP) was employed to perform the computations in this research [32], [33]. All the parameters are evaluated using the Generalized Gradient Approximation (GGA) [34] with Perdew-Burke-Ernzerhof (PBE) functional [35] in Density Functional Theory (DFT) [36]. To compute the interactions between electrons and ions, projector augmented wave (PAW) [37] is taken into consideration. The Heyd-Scuseria-Ernzerhof (HSE06) [38] hybrid functional is more accurate in describing the band structure of strongly correlated systems [39]. To get the correct band structure the HSE05 functional is used in this work. A convergence threshold of force was chosen as 0.001 eV/Å, and the cutoff energy is 500 eV. Up until the self-consistent total energy was changed to $10^{-8}$ eV/atom, the whole unit cell's ionic coordinates, shape, and volume were completely relaxed. The Brillouin zone integral was summed throughout the entire Brillouin zone using a K-mesh of Monkhorst–Pack 4×4×4 for F-based compound and 3×3×3 for Cl and Br-based double perovskites. Since all of the investigated crystal structures were fully optimized, the larger k-points and higher energy cutoff have little or no impact on the computed outcomes.

## Results and discussions:

### Structural Properties:

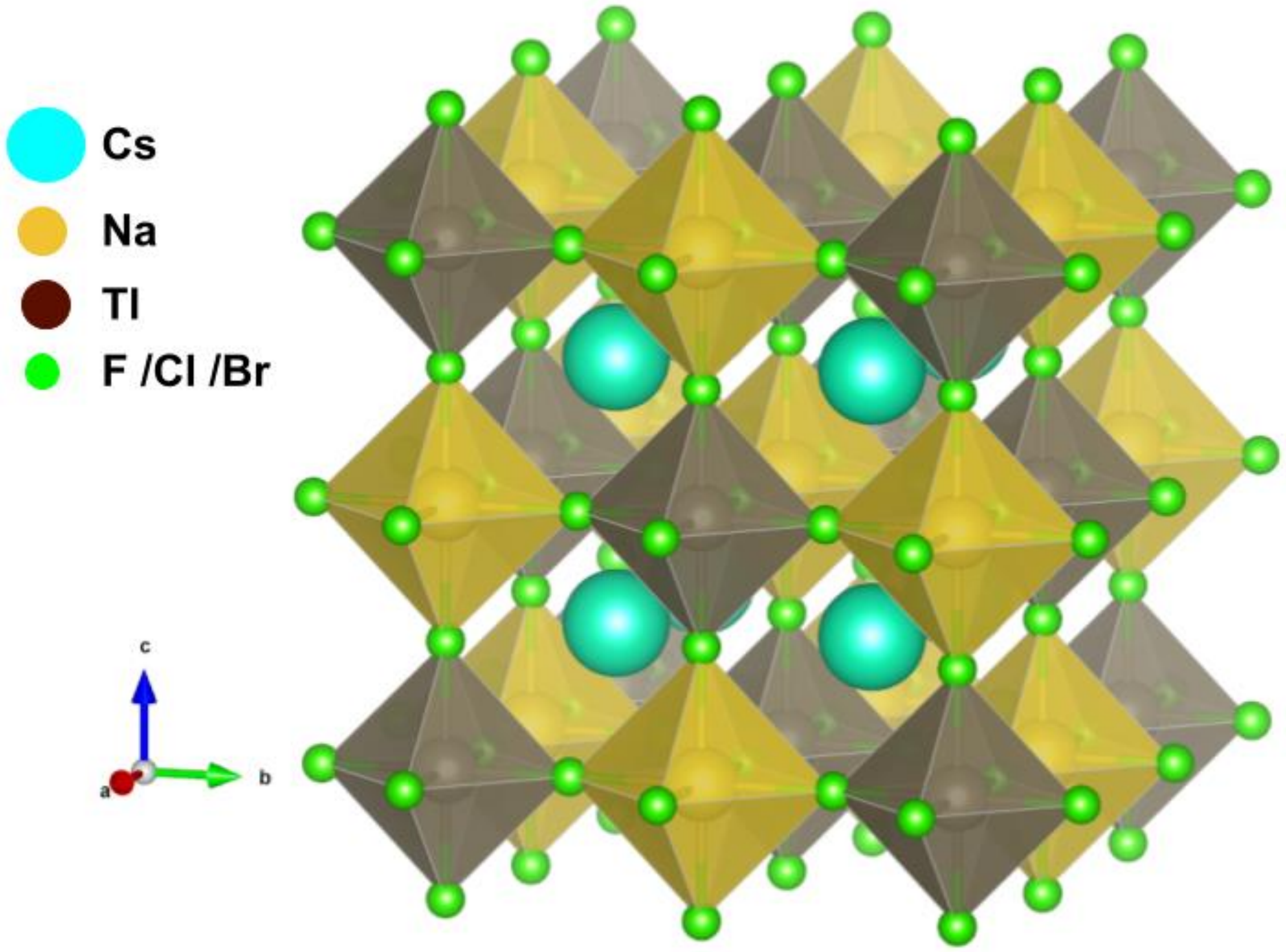

**Fig. 1: Geometrically optimized cubic crystal of $Cs_2NaTlX_6$ (X=F/Cl/Br) double halide perovskite.**

The face-centered cubic (rock salt) crystal structures of $A_2BB'X_6$ (A = Cs, B = Na, B' = Tl, X = F, Cl, Br) are presented in Fig.1, which belongs to the space group Fm-3m (225).

The equilibrium lattice parameter $a_0$, total volume $V_0$, total energy $E_0$, the bulk modulus $B_0$, and its pressure derivative $B'_0$ using GGA-PBE functional are tabulated in Table-1. The lattice parameter, total cell volume, and pressure derivative of bulk modulus are found to increase in the order of F < Cl < Br based double perovskites. This can be attributed to the enlargement of atomic size in this order. On the other hand, the total energy and bulk modulus of the double perovskite decreases with the increase of the atomic size. The equilibrium lattice parameter $a_0$ (in Å) is compared with available experimental [40] and theoretical [41] data for the $CS_2NaTlCl_6$ compound and found to be in good agreement. To the best of our knowledge there is no experimental data available so far for $CS_2NaTlBr_6$ and $CS_2NaTlF_6$. The total energy of the unit cell for different cell volumes are calculated and fitted the energy per atom versus volume per atom curves in Fig.2 using the third-order Birch–Murnaghan equation [42].

$$P(V) = \frac{3}{2}B_0[(V_0/V)^{\frac{7}{3}} - (V_0/V)^{\frac{5}{3}}]\{1 + \frac{3}{4}(B'_0 - 4)[(V_0/V)^{\frac{2}{3}} - 1]\} \qquad (1)$$

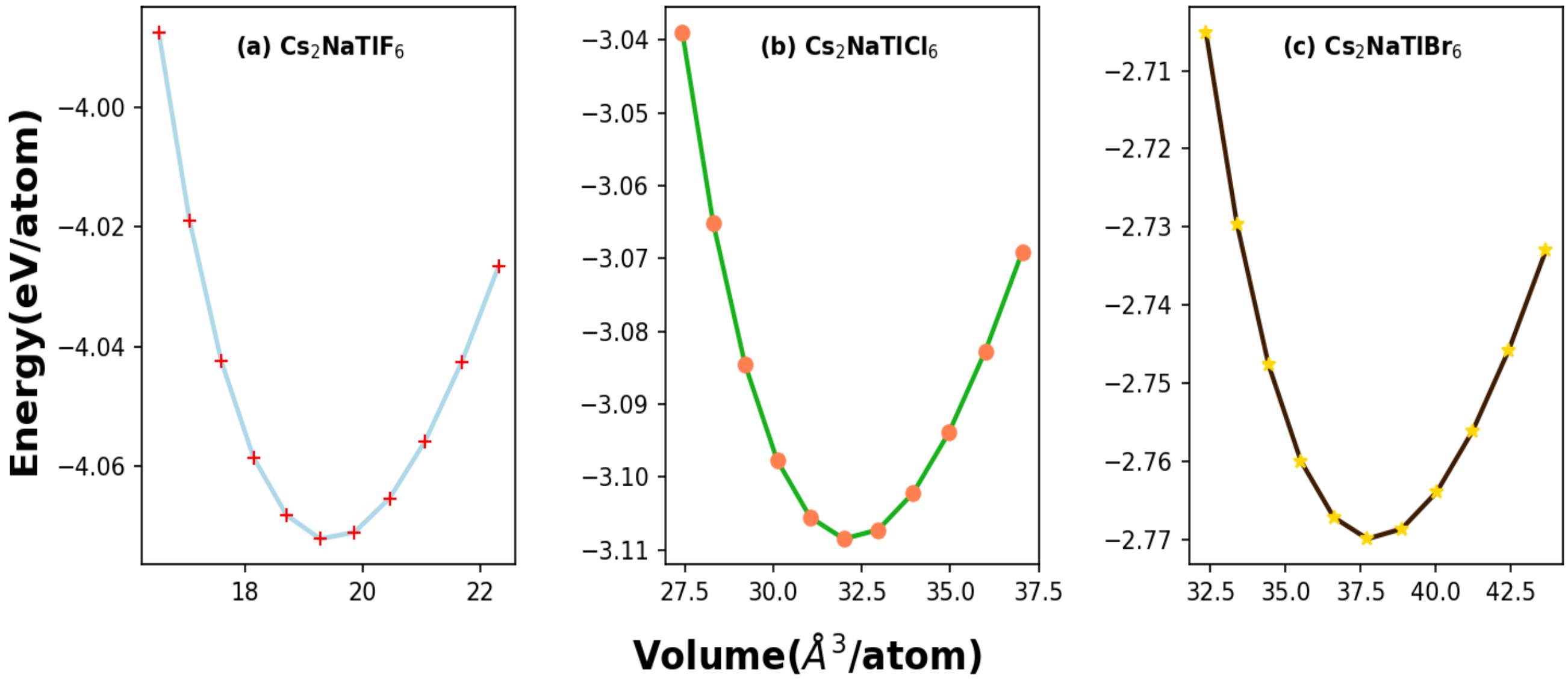

**Fig. 2: Computed energy variations by volume of the crystals for the studied double halide perovskite.**

The stability of perovskite is usually estimated by the tolerance factor $t = (RA + RX)/\sqrt{2}(RB + RX)$ of the Goldschmidt rule, where RA and RX are the radii of the ion positioned at A and B site (Na and Tl in this case) respectively, and RB is the average ionic radius of the elements at B and B' sites. A structure will be stable if the tolerance factor is within the range of $0.81 < t < 1.1$ [43]. The tolerance factors for the $Cs_2NaTlF_6$, $Cs_2NaTlCl_6$, and $Cs_2NaTlBr_6$ compounds are found to be 0.89, 0.85, and 0.85 respectively as presented in Table-1. All values of t are within the range satisfying the stability criterion of the Goldschmidt tolerance factor, which indicates that all the studied compounds are stable.

| Table-1: Computed structural parameters for the studied $Cs_2NaTlX_6$ double halide perovskites. | | | | | | | |
|---|---|---|---|---|---|---|---|
| **Compound** | **$a_0$ (Å)** | **$V_0$ (Å$^3$)** | **$E_0$ (eV)** | **$B_0$ (GPa)** | **$B_0$'** | **Tolerance factor, t** | **Formation Energy,** $\Delta E_f$ |
| **$CS_2NaTlF_6$** | 9.17, 8.995[44] | 776.86 | -162.89 | 45.31 | 5.09 | 0.89 | -3.047 |
| **$CS_2NaTlCl_6$** | 10.85, 10.62[45], 10.58[41] | 1286.76 | -124.34 | 23.00 | 5.19 | 0.85 | -1.484 |
| **$CS_2NaTlBr_6$** | 11.47, 11.13[46] | 1516.71 | -110.80 | 18.21 22.35[46] | 5.27 | 0.85 | -1.477 |

The formation energy ($\Delta E_f$) of a compound is a measure of thermodynamic stability of a material and is estimated by the following equation[47]:

$$\Delta E_f = [E_{total} - \sum_a nE_a]/N \qquad (2)$$

where N = Number of atoms in the unit cell. For example, the formation energy for the $Cs_2NaTlX_6$ (X = F, Cl, Br) compounds is given by the following formula

$$\Delta E_f(Cs_2NaTlX_6) = E_{Cs_2NaTlX_6} - 8E_{Cs} - 4E_{Na} - 4E_{Tl} - 24E_X)/40 \qquad (3)$$

A negative value of the formation energy indicates that the formation of the compound is exothermic and therefore thermodynamically stable, as the energy released during the formation process is greater than the energy required to form the compound. Our calculated values of formation energies of all the Cs2 NaTlX6 (X= F, Cl, Br) compounds are presented in Table-1, and it is found that the negative values of formation energy of all the studied compounds indicates that the compounds are thermodynamically stable and exothermic. The formation energy can also be related to the Gibbs free energy of formation, which is a measure of the spontaneity of the reaction. In this case, the negative values of the formation energy indicate that these compounds can be formed spontaneously under standard conditions. This is in accordance with the thermodynamic principle of decreasing Gibbs free energy for spontaneous reactions [48]. The stability of these compounds can also be understood from the perspective of the crystal structure. The stability of a crystal structure is dependent on the strength of the chemical bonds between the atoms, which is a direct result of the electron-electron interactions in the system [49], [50]. As the formation energy of these compounds is negative, it can be inferred that the electron-electron interactions in these compounds are favorable and lead to a stable crystal structure.

**Electronic Properties:**

Metals, semimetals, semiconductors, and insulators can be distinguished based on their electronic properties, and the computed electronic band structure can be used to recommend the materials under consideration for specific industrial applications.

The electronic band structures for $Cs_2NaTlX_6$ are calculated using GGA-PBE and hybrid (HSE06) functionals, which are shown in Fig. 3 and Fig. 4 respectively, and the band gap values are reported in Table-2. We have got direct band gap values of 3.58, 1.75, and 0.80 eV for $Cs_2NaTlF_6$, $Cs_2NaTlCl_6$, and $Cs_2NaTlBr_6$ respectively, executing GGA-PBE functional at the center of the Brillouin zone (Γ point). It is clearly observed that the values of band gap or semiconducting nature decrease with the increasing the radius of the halide ions from F to Br. This tendency might be caused by the decrease in the electronegativity difference between B'-site elements like Tl (1.62) and halide ions like F, Cl, and Br (3.98, 3.16, and 2.96 respectively), which strengthens the covalent bond between Tl and X element. This in turns may then be push the Tl-X bonding orbital and increase the energy of the VBM and reduces the bandgap of the materials. The direct band-gap semiconductors are expected to be suitable for photovoltaics and optoelectronic devices applications [51]–[54].

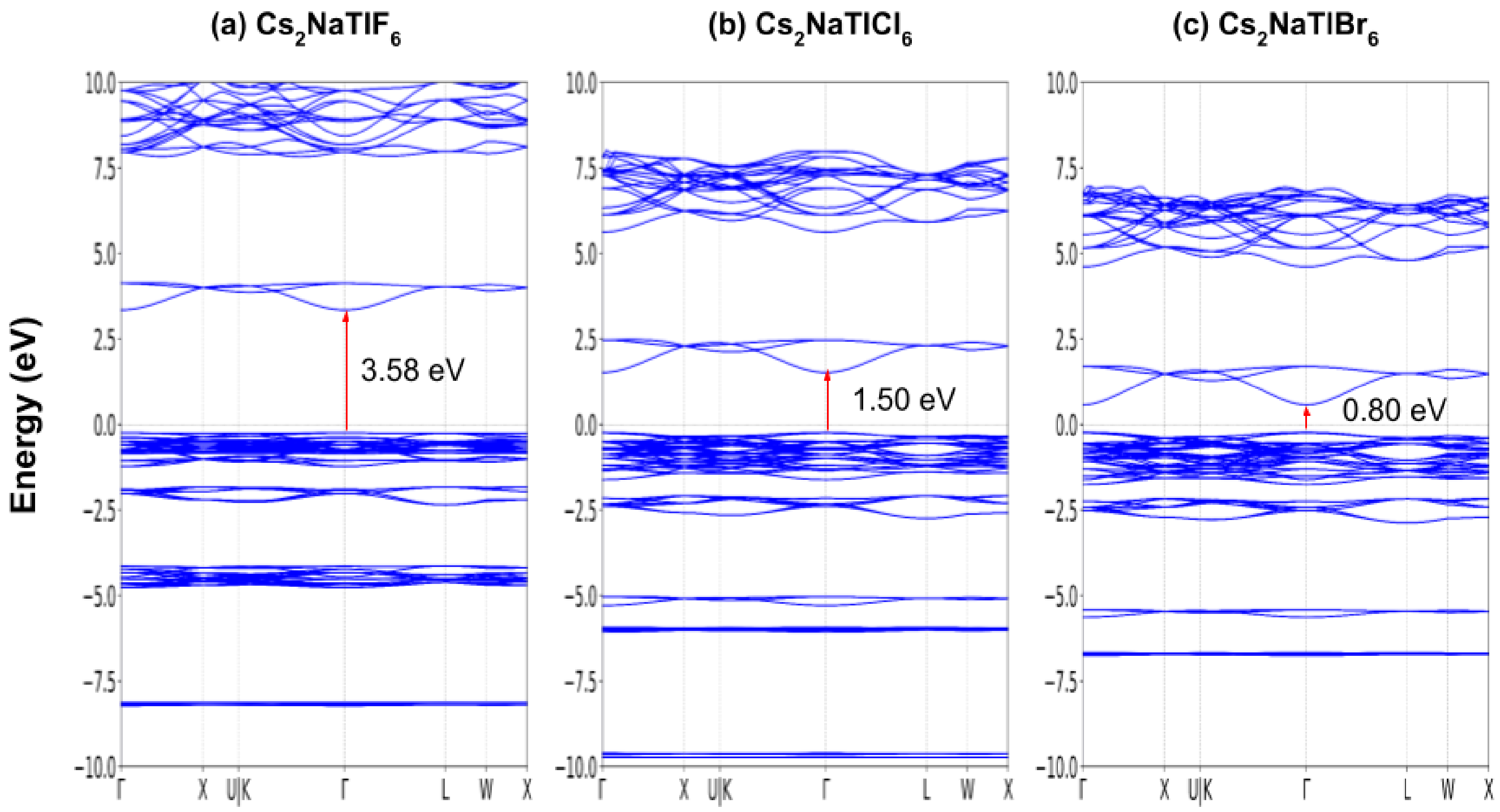

**Fig. 3: Energy band structure (GGA-PBE) for (a) $Cs_2NaTlF_6$, (b) $Cs_2NaTlCl_6$ and (c) $Cs_2NaTlBr_6$ double halide perovskites.**

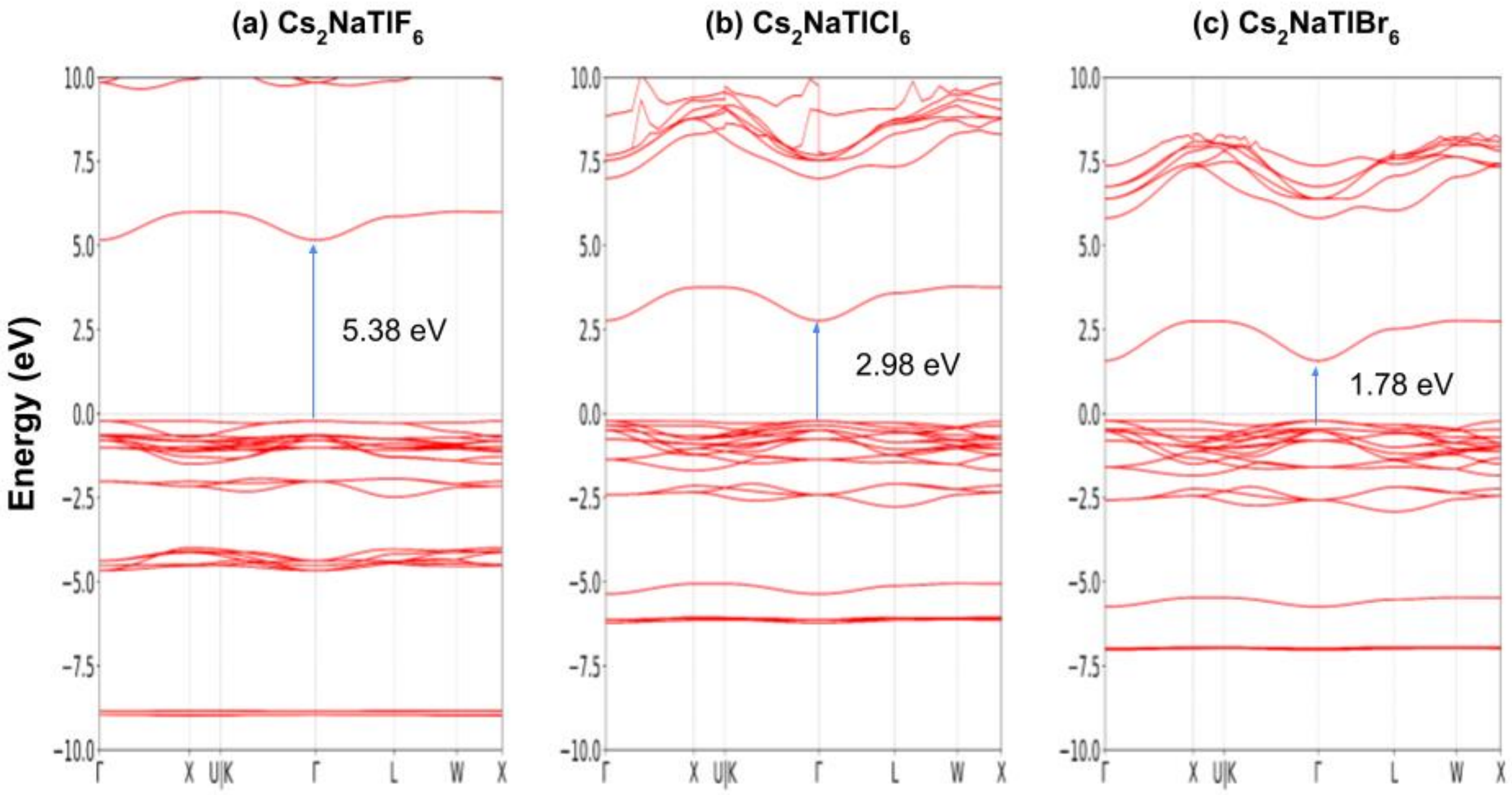

**Fig. 4: Calculated band gaps (HSE06) for (a) $Cs_2NaTlF_6$, (b) $Cs_2NaTlCl_6$ and (c) $Cs_2NaTlBr_6$ double halide perovskites**

Since we were aware that GGA-PBE functional underestimates the bandgap values, we decided to calculate the bandgap using hybrid functional (HSE06), as this functional compute the band gap value more accurately. It is important to note that the well-known self-interaction effect causes the very noticeable difference in bandgap between the value as calculated by using GGA-PBE and HSE06 functional. The values of bandgap for $Cs_2NaTlF_6$, $Cs_2NaTlCl_6$, and $Cs_2NaTlBr_6$ are calculated to be 5.38 eV, 2.98 eV, and 1.78 eV respectively within the HSE06 functional. It is clear from the bandgap value that the F-based compound exhibits an insulating nature, and the other two compounds ($Cs_2NaTlCl_6$ and $Cs_2NaTlBr_6$) behave like semiconductors. The values of the bandgap are comparatively greater in HES06 calculation than in the GGA-PBE calculations as the conduction bands are shifting towards higher energy value.

**Table-2: Estimated band gap and effective mass values for the studied double halide $Cs_2NaTlX_6$ (X=F, Cl, Br) perovskites.**

| **Compound** | **Bandgap (eV)** | | **$m_e$*** | | |
|---|---|---|---|---|---|
| | **GGA-PBE** | **HSE06** | **Γ→K,** | **Γ→L,** | **Γ→X** |
| $CS_2NaTlF_6$ | 3.58 | 5.38 | 0.762 | 0.762 | 0.764 |
| $CS_2NaTlCl_6$ | 1.75 | 2.98 | 0.437 | 0.437 | 0.437 |
| $CS_2NaTlBr_6$ | 0.80 | 1.78,<br>1.82 [46]<br>mbj+SOC | 0.282 | 0.283 | 0.283 |

The energy dispersion curve is fitted around the bottom of the conduction bands and the top of the valence bands to determine the effective mass ($m^*$) of electrons, which can be determined by the relation (4),

$$m^* = \hbar^2 \left[\frac{\partial^2 \varepsilon(k)}{\partial k^2}\right]^{-1} \quad (4)$$

where k is the wave vector and ε(k) is the band-edge eigenvalues.

We calculated the effective mass of electrons $m_e$* along three different paths (Γ→K, Γ→L, Γ→X) in k space and present in the Table-2. High carrier mobility is essential for the high performance of optoelectronic devices. Br-based compound possesses smaller electron effective masses than the other two materials which means that this compound realized the highest carrier mobility among the studied materials. In general, the carrier mobility is directly related to the electronic properties of the material, such as the band structure and the nature of the charge carriers. In the case of the halide double perovskites Cs2NaTlX6 (X= F, Cl, Br), the order of carrier mobility is Br>Cl>F-based compounds. This can be understood by considering the nature of the halogen element (X) in each compound. The halogen element acts as a dopant in these materials, introducing impurities that act as additional charge carriers. The larger the halogen element, the more impurities are introduced and the greater the number of charge carriers available to move through the material. Considering Br-based compounds, the larger atomic size of bromine leads to the introduction of more impurities and hence more charge carriers, resulting in higher carrier mobility compared to Cl-based compounds. This can be attributed to the larger ionic radius of the Br atom, leading to a more loosely packed crystal lattice and thus, less scattering of carriers. The band structure calculations of these compounds also showed that the Br-based compound had a higher density of states near the Fermi level, indicating a higher carrier concentration and thus, a higher mobility. These results suggest that the Br-based compound may have potential for use in electronic devices such as transistors and solar cells, where high carrier mobility is desired. Similarly, Cl-based compounds have higher carrier mobility than F-based compounds.

It should be noted that this conclusion is based on theoretical calculations using density functional theory (DFT) and the GGA-PBE method, and further experimental studies are needed to confirm the predicted order of carrier mobility in these materials. Additionally, other factors such as structural defects, crystal quality, temperature, etc. also affect the carrier mobility which is important to consider.

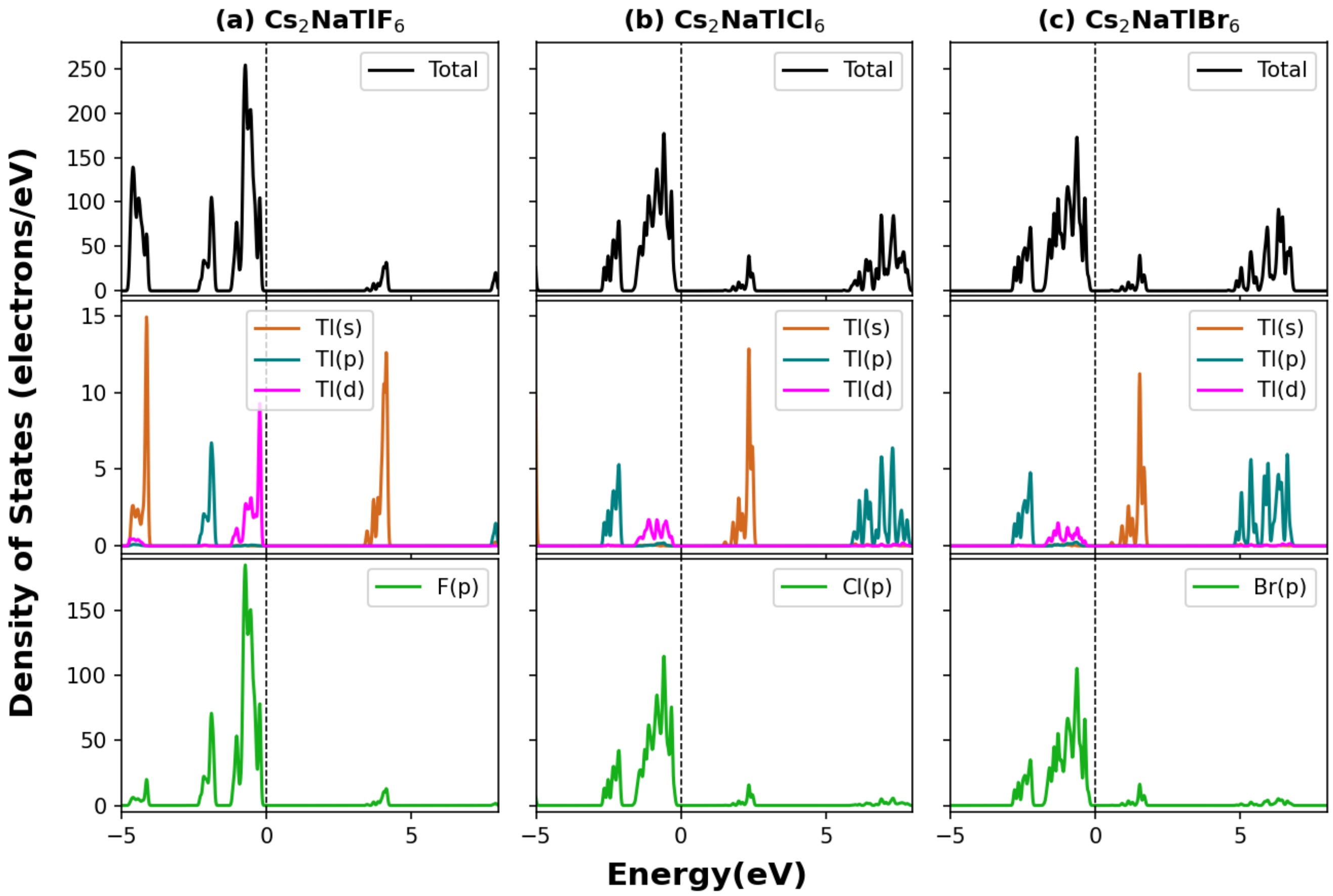

**Fig. 5: DFT calculated electronic properties with TDOS and PDOS of (a) $Cs_2NaTlF_6$, (b) $Cs_2NaTlCl_6$ and (c) $Cs_2NaTlBr_6$ double halide perovskites.**

The total density of states (TDOS) and partial density of states (PDOS) of materials provides a more detailed insight about the orbital contributions to the VBs and CBs in the electronic structure. Fig. 5 shows the computed TDOS and PDOS for $Cs_2NaTlF_6$, $Cs_2NaTlCl_6$ and $Cs_2NaTlBr_6$ with energies ranging from -5 eV to 8 eV. It is clear from the Fig. 5 that the d orbital of the Tl atom (5d) and the p orbital of the halogen atom (2p for F, 3p for Cl and 4p for Br) contribute significantly to the top of the VB, whereas the wide 5S orbital of Tl atom and p orbital of the halogen atoms are the major contributors to the bottom of the CB. A small contribution of Tl 3d states to the top of the VB is observed, but it decreases gradually from F to Br element-based materials. The presence of Tl 5s and Tl 5p in Cs2NaTlF6, Cs2NaTlCl6, and Cs2NaTlBr6 suggests that substitution of the F ion with Cl and subsequently with Br does not significantly impact the valence states of the Tl atom. However, the conduction states are significantly shifted to lower energy levels.

The perovskites Cs2NaTlCl6 and Cs2NaTlBr6 possess semiconducting nature and direct band gap values, rendering them suitable for various optoelectronic applications. The ability to tune band structures and bandgaps by altering the atoms at the X-sites in double halide perovskite materials containing NaTl has led to their prominence as photoelectric conversion materials.

**Optical Properties:**

Optical properties are the most important in determining a material's suitability for optoelectronic and photovoltaic applications since they provide an insight on how well a material interacts with light. To comprehend a material's electronic configuration and assess its feasibility for optoelectronic applications, understanding the optical properties of a material is vital. This study examines the detailed optical characteristics of the considered double-halide perovskite materials, including photo- absorption coefficient α(ω), reflectivity (R), refractive index n(ω), dielectric constants (ε), optical conductivity (σ), energy loss function (L), and extinction coefficient (k) corresponding to photon energy (E eV) of electromagnetic spectrum. To analyze the optical properties of the $Cs_2NaTlX_6$ (X = F, Cl, Br) double perovskite, density functional theory (DFT) employed first-principles calculations performed using Pseudo-potential and GGA-PBE approximation in VASP. The profile of the absorption coefficient α(ω) is calculated by using equation (5) is essential for determining the light harvesting efficiency of the perovskite materials [55]:

$$\alpha(\omega) = \frac{\sqrt{(2\omega)}}{c}\sqrt{[(K) - \varepsilon_1(\omega)]} \qquad (5)$$

$$\text{where } K = \sqrt{\left(\varepsilon_1^2(\omega) + \varepsilon_2^2(\omega)\right)} \qquad (5a)$$

$\varepsilon_1$, $\varepsilon_2$ respectively denotes the real and imaginary part of the frequency dependent dielectric function. To understand the fundamentals of photogenerated carriers, polarization nature and the exciton binding energy; dielectric constant for the double perovskites are calculated. Applying the equation (6), complex frequency dependent dielectric functions ε(ω) is determined as [56]:

$$\varepsilon(\omega) = \varepsilon_1(\omega) + i\varepsilon_2(\omega) \qquad (6)$$

where $\varepsilon_1$, $\varepsilon_2$ respectively denotes the real and imaginary part of dielectric function. Electrical susceptibility and complex susceptibility can be determined using the dielectric parameters following relation [57]:

$$\chi(\omega) = \varepsilon_r(\omega) + 1 = i\omega P(\omega) \qquad (6a)$$

However, other necessary optical parameters are refractive index n(ω), optical conductivity σ(ω), energy loss function L(ω) and extinction coefficient k(ω) are calculated using the following relations [58]:

$$n(\omega) = \frac{1}{\sqrt{2}}\sqrt{[(K) + \varepsilon_1(\omega)]} \qquad (7)$$

$$\sigma(\omega) = \frac{\omega\varepsilon_2}{4\pi} \qquad (8)$$

$$L(\omega) = \frac{\varepsilon_2(\omega)}{K^2} \qquad (9)$$

$$k(\omega) = \frac{1}{\sqrt{2}}\sqrt{[(K) - \varepsilon_1(\omega)]} \qquad (10)$$

In Fig (6)-(8), DFT extracted optical parameters are represented for 0- 35 eV of photon energy. The photo-absorption coefficient (α) describes how much light penetrates a medium before being absorbed. Fig. 6 displays the photon energy-dependent absorption characteristics of the considered $Cs_2NaTlX_6$ double perovskite compounds. The absorption profile can be estimated using equation (5) by taking into account the values of the dielectric functions, which proportionally affect the variation of the absorption profile. Noteworthy, optical band gaps found from DFT calculations using PBE functional are underestimated, but the absorption profile for the considered double perovskites quite agrees with the fundamental energy bandgap as displayed in Fig. 3. Within the photon energy range (1.5-3.1 eV) of visible spectrum, the absorption profile for the $Cs_2NaTlX_6$ halide double perovskites steadily increase with increment in energy. However, for all the compositions, three major peaks are observed in the whole energy range where another minor peak are observed near about 30 eV energy as exhibits in Fig. 6. As shown in Fig. 6, the maximum absorption peak is seen for all of the compositions at 13 eV, and $Cs_2NaTlF_6$ exhibited the highest amplitude among all the three. The second highest absorption peak is seen in the region of 7 eV to 10 eV for the studied perovskites. In the visible range of photon energy, $Cs_2NaTlBr_6$ double perovskite leads to others having a predisposition to exhibit more dominant behavior. However, the first peak for all of the materials is found in the low energy region between 2.3 eV and 5 eV which is close to the HSE06 functional bandgap as listed in Table 2. Substitution in X site by F, Cl, and Br leads to magnitude variation of the optical absorption for $Cs_2NaTlX_6$ double halide perovskites. Fig. 6 exhibits that F containing perovskite demonstrate highest peak among the studied materials with a wider absorption profile. In the visible range of photon energy, Br contained perovskite shows the dominance over others. As the visible range of the electromagnetic spectrum is where the sun radiates the most of its energy, the fact that the material absorbs more light in the lower photon energy implies that it is suitable for solar cell applications. Changes in X site from F to Cl and Br leads to a left shift to lower energies in the absorption profile for the studied materials which clearly demonstrate the effect of ionic radius changes of halogen ions in the structure. As a consequence of the lowest exciton binding energy leading to the largest carrier mobility, perovskites containing Br are more efficient at the photoelectric conversion utilizing visible light than those without Br.

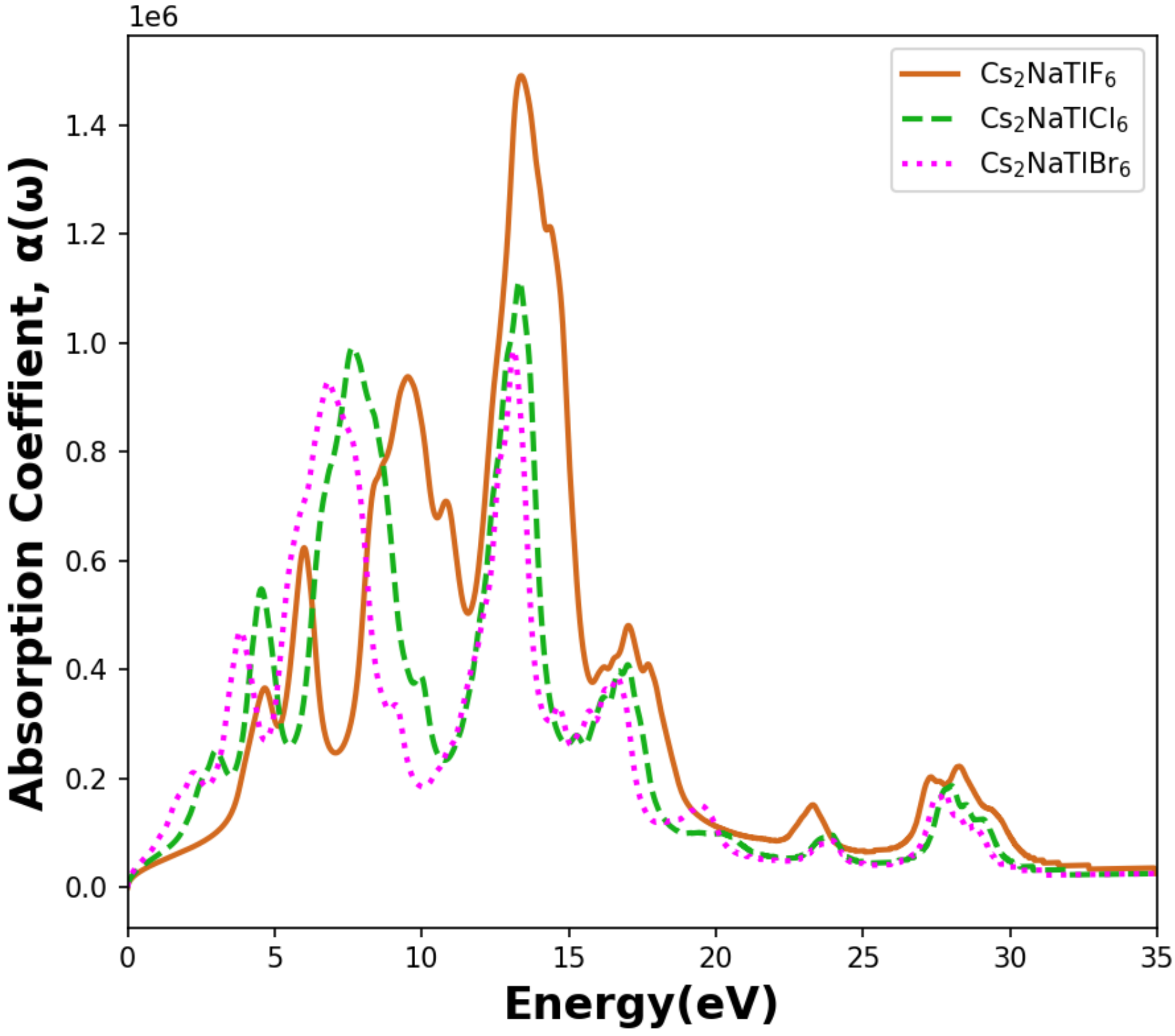

**Fig. 6: Photo-absorption spectra for Cs2NaTiX6 double perovskites.**

The dielectric function ε(ω), which explains how a material interacts with an external electromagnetic field, is the most important optical characteristic where all of a material's optical aspects are interrelated. Eq. (6) represents the complex dielectric function with real ($\varepsilon_1(\omega)$) and imaginary ($\varepsilon_2(\omega)$) parts' combination which is used to measure the dielectric nature of the materials. Since, optical properties of a crystal depends on the structure, hence in turn it depends on the band structures. The inter-band transitions significantly contribute to ε(ω) in semiconductor materials where intra-band contribute to the complex dielectric function for metal type materials [58]. The desired dielectric responses for the studied materials are illustrated in Fig. 7 in the photon energy range of 0-35 eV. From Fig 7(a), it is observed that $\varepsilon_1(\omega)$ exhibits its maximum peak with first harmonic after a gradual rise at 1.35 eV, 2.37 eV, 3.98 eV for F, Cl, Br based perovskite materials respectively.

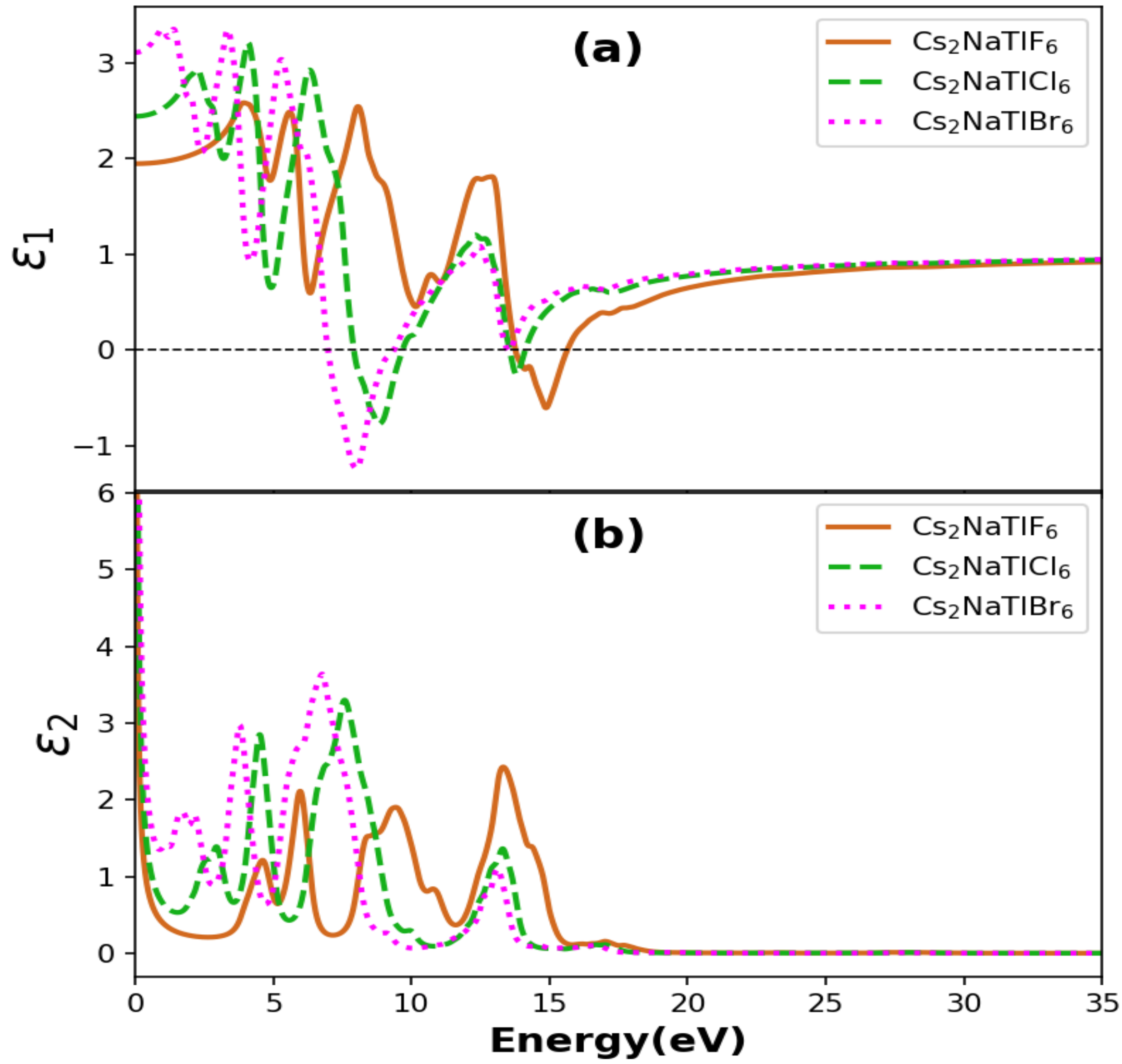

**Fig. 7: Complex Dielectric functions for Cs2NaTiX6 double perovskites (a) Real part and (b) Imaginary part.**

The evaluated perovskite materials' apparent dielectric variation suggests that the materials are highly transmittable in the high energy area, however low transmittance is evident for the low energies. A material's sensitivity to incident light energy determines the optoelectronic device efficacy where dielectric constant is the crucial parameter to describe. The higher values of dielectric constant at a lower rate of charge recombination have a substantial impact on the performance of optoelectronic devices. It is evident, based on both the imaginary and real portions of the dielectric function, that Cs2NaTiBr6 perovskite has a higher amplitude in the visible spectrum than other two materials. As observed from Fig. 7, $\varepsilon_1(0)$ is increased with changing F by Cl and Br indicates higher polarizability of larger halogen atoms in the studied compounds [59].

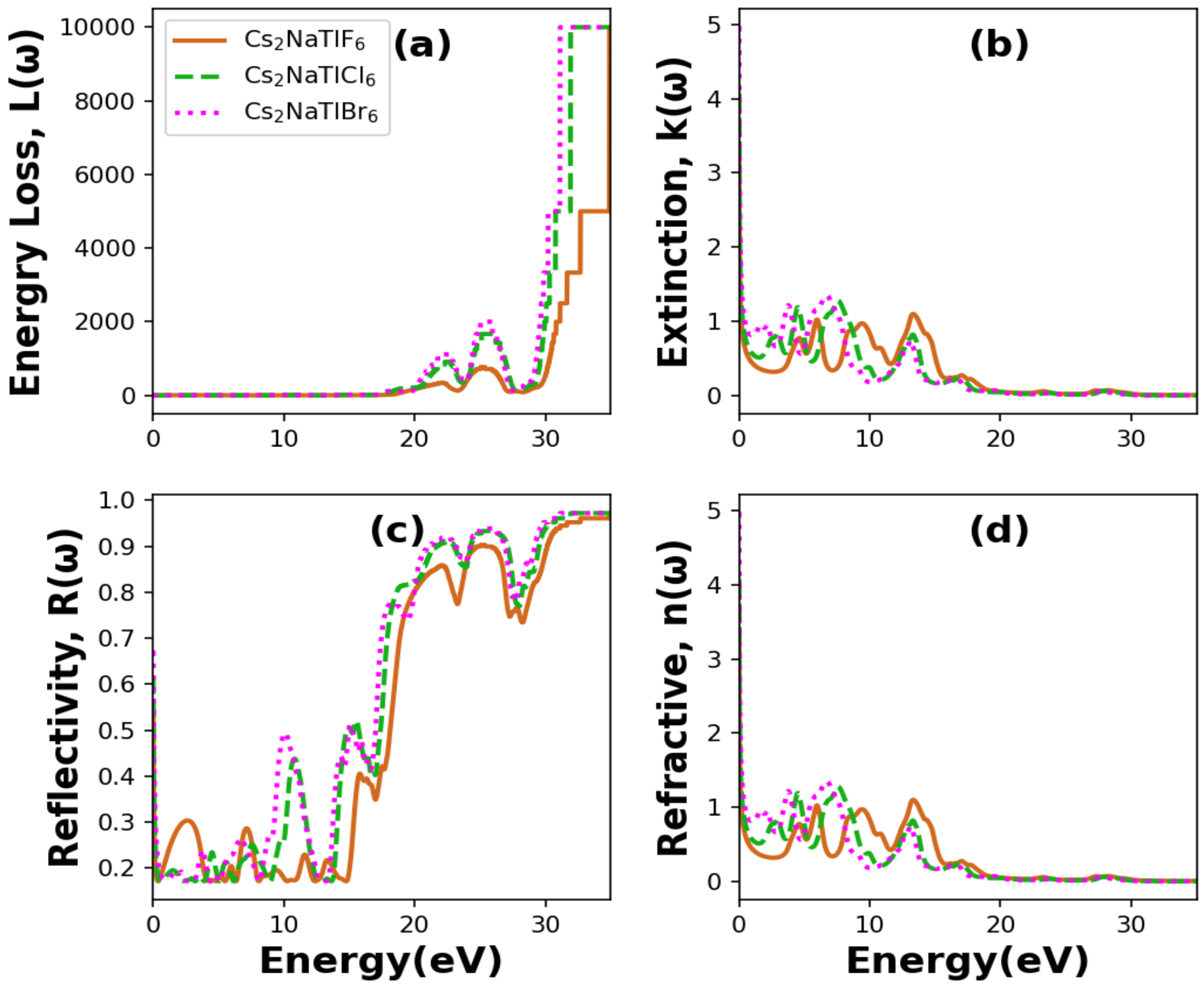

**Fig. 8: The evaluated energy loss function, reflectivity, refractive index, and extinction coefficient for the Cs2NaTlX6 double halide perovskites.**

Another important parameter is the energy loss function to assess the optical properties, which describes electrons' fast-moving effect on energy loss in the material. In Fig. 8(a), energy loss function with respect to the photon energy (0-35 eV) is portrayed for the studied perovskites. It is shown that electron energy loss increases in higher energy region whereas this value is zero in the visible energy region and up to the photon energy of 17 eV. The energy loss functions indicate that studied perovskites starts to have finite energy loss when energy to ionize atom range starts beyond the binding energy of a tightly bound molecule. Cs2NaTiBr6 double halide perovskite shows dominance among others in the higher region to increase the loss sensitivity with the energy as shown in Fig. 8(a).

To understand the surface nature of perovskite materials, reflectivity, another one optical parameter needs to be examined. Reflectivity describes the interaction of light with the surface of a material, specifically the ratio of light energy reflected off the surface to light energy incident on the surface. The DFT simulated reflectivity spectra for the three double halide perovskites are represented in Fig. 8(b). It is observed from Fig. 8(b) that in the lower energy region (up to 15 eV) all three double halide perovskites exhibit lower reflectivity. All of the $Cs_2NaTlX_6$ perovskites provides higher magnitude of reflectivity in the higher photon energy region identically that indicates significant absorptivity and/or transmission in the materials [60]. An exceptional pick is observed for zero energy scale for Br and Cl contained materials. However, in the visible light energy region, F based double perovskite $Cs_2NaTlF_6$ reflect more energy than Cl and Br based perovskites and maximum reflectivity of 0.3 is observed where 0.24 is observed for both of the Cl and Br based materials. The observed low reflectivity in the visible energy region indicates about the materials’ suitability in solar cell applications.

The degree of absorption loss, and hence the device efficiency, is determined by the extinction coefficient k(w), another crucial optical indication for predicting materials' potential for photovoltaic and other optoelectronic device applications [61]. Fig. 8(b) shows the variation of DFT calculated extinction coefficient in respect to photon energy. The extinction coefficients rise toward higher values within the visible spectrum with changing the halogen ions by larger ionic radius from F to Br. Comparing with F and Cl, Br shows its higher values of 0.9 in the visible energy region. A maximum peak of 1.35 is showed by Cs2NaTiBr6 perovskite in the whole spectrum at 7.3 eV. However, dominance of Br shrinks to the higher energy whereas Cs2NaTiF6 double perovskite shows peak of higher magnitude in the high energy region as shown in Fig. 8(b). the Cs2NaTiCl6 behaves same as Cs2NaTiBr6 in the whole spectrum of the extinction coefficients except it is slightly lower in magnitude.

**Table-3: The calculated values of static dielectric constant $\varepsilon_1(0)$, static refractive index n(0), static reflectivity R(0) and susceptibility ($\chi$) for the studied double halide perovskites.**

| Compound | $\varepsilon_1(0)$ | n(0) | R(0) | $\chi$ |
|---|---|---|---|---|
| **$CS_2NaTlF_6$** | 1.93 | 1.56 | 0.66 | 2.93 |
| **$CS_2NaTlCl_6$** | 2.48 | 4.05 | 0.61 | 3.48 |
| **$CS_2NaTlBr_6$** | 3.13 | 5.00 | 0.54 | 4.13 |

To comprehend how light travels through materials, understanding of optical refraction is essential. The light will move closer to its normal direction as the refractive index goes up. The calculated refractive index of the double-halide $Cs_2NaTlX_6$ (X=F, Cl, Br) perovskites is shown as a function of photon energy up to 35 eV in Fig. 8(d). Like extinction coefficients' behavior, all of the perovskites have a refractive index that is qualitatively identical as observed so far. It has been found that the static refractive index ranges from a high value of 5.0 for $Cs_2NaTlBr_6$ at E=0 eV to a low value of 1.0 for $Cs_2NaTlF_6$. The refractive index of perovskites exhibits very identical patterns regardless of whether Cl or Br is present in the X sites, with the exception that the peak value for Cs2NaTiBr6 is greater than others.

The results of optical properties for the studied compounds exhibit that Cl and Br contained CsNaTl double halide perovskites have higher absorption coefficient, larger dielectric constant, and higher value of refractive index than F contained perovskite, which are favorable properties for solar cell application. Also, it suggests that the perovskites may be useful for energy harvesting applications, such as in photovoltaic devices. To enhance light-matter interactions in such materials through accumulating optical behaviors and energies utilizing coherent electron confinement, surface plasmon principles can be engineered. The carrier density of semiconductors can enhance the plasmonic properties this in turns can play a vital role in such crystals for photovoltaic applications. Multiple scattering effects appear due to metallic structures results enhancement in photon's optical paths length, which leads to light harvesting efficiency as well as charge-carrier separation efficiency of the compound, plasmonic metals can be incorporated in the perovskite crystals for higher solar cell efficiency [62]. Moreover, with interface layer of perovskite cell, Epsilon Near Zero (ENZ) transition layer can be placed to enhance such halide perovskite solar cell's efficiency considerably [63].

**Mechanical Properties:**

For any crystal-solid, elastic tensor give a comprehensive insight about the material's mechanical stability. In this study, finite strain theory was employed to calculate the values for elastic parameters of the double halide $Cs_2NaTiX_6$ (X = F, Cl, Br) perovskites. Table 4 summarizes the findings of a stress-strain analysis performed on three different elastic moduli $C_{ij}$ ($C_{11}$, $C_{12}$, and $C_{44}$) to evaluate the mechanical behaviors of the materials in consideration. Having cubic symmetry in perovskite crystal should be followed by the Born-stability criterion as $C_{11}>0$, $C_{44}>0$, $C_{11}-C_{12}>0$, and $C_{11}+2C_{12}>0$ [60]. It is evident from Table 4 that the studied perovskites satisfied well with the criterions, indicates that the $Cs_2NaTiX_6$ (X = F, Cl, Br) perovskites are mechanically stable. In addition, the double halide perovskites under study satisfy the cubic crystal stability condition: $C_{12} < B < C_{11}$ of the elastic tensor. In general, $2C_{44} = C_{11}-C_{12}$ should persist for elastically isotropic cubic crystals [64]. Table 4 shows that C11 is much greater than C12 and C14, indicating high resistance to flex in the [100] ([010] or [001]) direction but low resistance to shear deformation. The octahedral distortion is account for this difference, since it suggests a stronger coupling between the a and c crystal direction than the a and b and b and c directions. To determine materials' brittleness and ductility, the Cauchy pressure (C12-C44) is an essential parameter value to be calculated. A metallic bond is indicated by a positive Cauchy pressure value, whereas a covalent bond is indicated by a negative value. The studied double perovskite materials exhibit positive values as observed from the retrieved Cauchy pressure values listed in Table 4. However, considering polycrystalline materials, Voigt-Reuss-Hill approximations are utilized using the single-crystal elastic constants in this study to evaluate other mechanical parameters for the double halide perovskites with the following relations [56], [64], [65]:

Bulk modulus: B = $\frac{B_v+B_R}{2} = \frac{C_{11}+2C_{12}}{3}$ (11)

Shear modulus: S = $\frac{G_v+G_R}{2} = \frac{C_{11}-C_{12}}{2}$ (12)

Young's modulus: Y = $\frac{9BG}{(3B+G)}$ (13)

Poisson's ration: v = $\frac{3B-2G}{2(3B+G)}$ (14)

Universal Anisotropy Index (UAI): $A^U = 5\frac{G_v}{G_R} + \frac{B_v}{B_2} - 6$ (15)

Accordingly, to get a deep understanding about materials' rigidity bulk modulus (B) is an essential parameter that we have calculated and listed in Table 4. The computed values of 35.77, 18.45, and 14.61 for $Cs_2NaTlF_6$, $Cs_2NaTlCl_6$ , and $Cs_2NaTlBr_6$ perovskite respectively reveal that all three materials exhibit flexibility and softness, which is an indication that the studied double halide perovskite materials are suitable for use in flexible and shape dependent optoelectronic devices. Changes in halide ions from F to Cl result to a lowering of the value of the bulk modulus [61]. A The longitudinal stress resistance capability of a material is determined by its value of Young's modulus E [66]. It is clear from Table 4 that the value of Young's modulus rises as the Bulk modulus rises. Changing in X site from F to Cl results to decrease in the magnitude of Young's modulus value as presented in Table 4.

| Table. 4: Evaluated elastic parameters for $Cs_2NaTlX_6$ (X=F, Cl, Br) double perovskites. | | | |
|---|---|---|---|
| **Elastic Parameters** | $Cs_2NaTlF_6$ | $Cs_2NaTlCl_6$ | $Cs_2NaTlBr_6$ |
| **C11** | 60.434 | 35.024 | 28.369, 45.09[46] |
| **C12** | 23.440 | 10.218 | 7.726, 10.98[46] |
| **C44** | 19.874 | 8.744 | 6.035, 7.82[46] |
| **$B_V$=$B_R$=$B_H$** | 35.771 | 18.487 | 14.607 |
| **$G_V$** | 19.323 | 10.208 | 7.750, 11.51[46] |
| **$G_R$** | 19.299 | 9.914 | 7.237, 9.98[46] |
| **$G_H$** | 19.311 | 10.061 | 7.493, 10.57[46] |
| **$E_V$** | 49.124 | 25.863 | 19.755, 29.43[46] |
| **$E_R$** | 49.073 | 25.231 | 18.634 |
| **$E_H$** | 49.099 | 25.548 | 19.197 |
| **Poisson's ratio, V** | 0.271 | 0.267 | 0.275 |
| **Poisson's ratio, R** | 0.271 | 0.273 | 0.287 |
| **Poisson's ratio, H** | 0.271 | 0.270 | 0.281, 0.28[46] |
| **$(B/G)_V$** | 1.851 | 1.811 | 1.885 |
| **$(B/G)_R$** | 1.854 | 1.865 | 2.018, 2.11[46] |
| **$(B/G)_H$** | 1.852 | 1.865 | 0.281, 0.28[46] |
| **Cauchy Pressure (GPa)** | 3.566 | 1.474 | 1.691, 3.16[46] |
| **$A^U$** | 0.006 | 0.148 | 0.354, 0.45[46] |
| **$V_l$** | 3429.163 | 2950.782 | 2397.061 |
| **$V_t$** | 1921.261 | 1657.105 | 1323.025 |
| **$V_a$** | 2138.368 | 1844.013 | 1474.279 |
| **Debye Temp.** | 237.456 | 172.937 | 130.870 |

For the studied $Cs_2NaTlX_6$ (X=F, Cl, Br) double perovskites, Pugh's ration (B/G) and Poisson ration (*v*) are assessed to determine the likely mode of failure (material's failure mechanism) and listed in Table 4. Distinguishing between ductility and brittleness of a material can be determined by its ductile-brittle transition threshold of these two parameters. The value of Pugh's ratio greater than 1.75 indicates the ductile nature of a material whereas for Poisson ratio it scales more than 0.26 [55] and makes them suitable for the fabrication of flexible devices such as thin films and geometry optimized optoelectronic devices. Considering the above criterion, the $Cs_2NaTlX_6$ (X=F, Cl, Br) perovskite materials indicate to be ductile as presented in the Table 4. Compound containing Br has a higher degree of ductility over the F and Cl contained materials.

Additionally, it has been observed that the anisotropy index of the double perovskite containing Br is higher than that of the materials containing F and Cl. This suggests that the double perovskite containing Br is sensitive to ambient conditions, although the magnitude of this sensitivity is negligible. The Debye temperature ($\Theta_D$) is also calculated and presented in Table 4. This thermodynamic parameter measures the possible temperature of a crystal to be generated owing to a single normal mode of vibration that impacts elasticity via the thermodynamic nature of a material. The thermal conductivity coefficient of a crystal is influenced by the mode of phonon vibration is directly correlated with the Debye temperature for phonon–phonon scattering. The dynamic behavior of dislocations (internal friction) in crystals on account of density of the thermal phonons changes essentially influenced by Debye temperature as the interaction of dislocations with conduction electrons increases significantly manifesting the possibility of superconducting transition as well as the quantum effects (quantum tunnelling and quantum oscillations) [67]–[69]. It is clear from the Table 4 that the Debye temperatures for these $Cs_2NaTlX_6$ (X=F, Cl, Br) are all much below 400 K, which indicates that $Cs_2NaTlX_6$ (X=F, Cl, Br) perovskites have a poor thermal conductivity and a low wave velocity. The mean velocity of sound is highest for F contained perovskite of 2138.368 $ms^{-1}$ whereas it decreases with Cl and Br substitution as seen from Table 4.

## Conclusion

In this study, we have performed a comprehensive theoretical analysis of the crystal structural, optoelectronic, and mechanical properties of double halide Cs2NaTlX6 (X = F, Cl, Br) utilizing density functional theory. We have found that the change in halogen site with a larger ionic radius (Br > Cl > F) leads to an energy band gap in the visible region, making these compounds more suitable for solar cell and optoelectronic applications. Structural investigations indicate that all the studied perovskite crystals are stable in the cubic phase, which is in good agreement with previously reported experimental values. Our calculations have shown that all three studied perovskites have a direct band gap nature in both PBE-GGA and HSE06 functionals. The Cs2NaTlF6 perovskite is an insulator with a direct band gap of 5.38 eV (HSE06), while Cs2NaTlCl6 and Cs2NaTlBr6 perovskites have semiconducting properties with direct band gap values of 2.98 eV and 1.78 eV, respectively. This indicates that Cs2NaTlCl6 and Cs2NaTlBr6 perovskites are particularly suitable for a wide range of optoelectronic applications. Our analysis of the optical properties of the studied perovskites suggests that Cs2NaTlCl6 and Cs2NaTlBr6 are suitable for solar cell applications, due to their high absorption coefficient, large dielectric constant, and high refractive index. Additionally, the ductile nature of these double halide perovskites makes them suitable for the fabrication of flexible devices such as thin films and shape dependent optoelectronic devices. Considering our findings, this study suggests that Cs2NaTlCl6 and Cs2NaTlBr6 perovskites may be beneficial for solar cells, optoelectronics, and energy harvesting applications.

## Author Contributions

M. M. Hasan designed the study and performed the calculations. M. M. Hasan, and N. Hasan analyzed the data and wrote the manuscript. A. Kabir supervised the investigations with guiding the computations. All authors reviewed the manuscript.

## Conflicts of Interest

The authors declare that they have no conflicts of interest.

## References

[1] A. Kojima, K. Teshima, Y. Shirai, and T. Miyasaka, "Organometal Halide Perovskites as Visible-Light Sensitizers for Photovoltaic Cells," *J Am Chem Soc*, vol. 131, no. 17, pp. 6050–6051, May 2009, doi: 10.1021/ja809598r.

[2] W. S. Yang *et al.*, "High-performance photovoltaic perovskite layers fabricated through intramolecular exchange," *Science (1979)*, vol. 348, no. 6240, pp. 1234–1237, Jun. 2015, doi: 10.1126/science.aaa9272.

[3] NREL, "Best research-cell efficiency chart.," 2020.

[4] M. D. McGehee, “Continuing to soar,” *Nat Mater*, vol. 13, no. 9, pp. 845–846, Sep. 2014, doi: 10.1038/nmat4050.

[5] S. de Wolf *et al.*, “Organometallic Halide Perovskites: Sharp Optical Absorption Edge and Its Relation to Photovoltaic Performance,” *J Phys Chem Lett*, vol. 5, no. 6, pp. 1035–1039, Mar. 2014, doi: 10.1021/jz500279b.

[6] A. Miyata *et al.*, “Direct measurement of the exciton binding energy and effective masses for charge carriers in organic–inorganic tri-halide perovskites,” *Nat Phys*, vol. 11, no. 7, pp. 582–587, Jul. 2015, doi: 10.1038/nphys3357.

[7] V. D’Innocenzo *et al.*, “Excitons versus free charges in organo-lead tri-halide perovskites,” *Nat Commun*, vol. 5, no. 1, p. 3586, May 2014, doi: 10.1038/ncomms4586.

[8] D. Ju *et al.*, “Tellurium-Based Double Perovskites $A_2TeX_6$ with Tunable Band Gap and Long Carrier Diffusion Length for Optoelectronic Applications,” *ACS Energy Lett*, vol. 4, no. 1, pp. 228–234, Jan. 2019, doi: 10.1021/acsenergylett.8b02113.

[9] G. Xing *et al.*, “Long-Range Balanced Electron- and Hole-Transport Lengths in Organic-Inorganic $CH_3NH_3PbI_3$,” *Science (1979)*, vol. 342, no. 6156, pp. 344–347, Oct. 2013, doi: 10.1126/science.1243167.

[10] C. Wehrenfennig, G. E. Eperon, M. B. Johnston, H. J. Snaith, and L. M. Herz, “High Charge Carrier Mobilities and Lifetimes in Organolead Trihalide Perovskites,” *Advanced Materials*, vol. 26, no. 10, pp. 1584–1589, Mar. 2014, doi: 10.1002/adma.201305172.

[11] S.-H. Turren-Cruz *et al.*, “Enhanced charge carrier mobility and lifetime suppress hysteresis and improve efficiency in planar perovskite solar cells,” *Energy Environ Sci*, vol. 11, no. 1, pp. 78–86, 2018, doi: 10.1039/C7EE02901B.

[12] A. Soni, K. C. Bhamu, and J. Sahariya, “Investigating effect of strain on electronic and optical properties of lead free double perovskite Cs2AgInCl6 solar cell compound: A first principle calculation,” *J Alloys Compd*, vol. 817, p. 152758, Mar. 2020, doi: 10.1016/j.jallcom.2019.152758.

[13] M. Lyu *et al.*, “Organic–inorganic bismuth (III)-based material: A lead-free, air-stable and solution-processable light-absorber beyond organolead perovskites,” *Nano Res*, vol. 9, no. 3, pp. 692–702, Mar. 2016, doi: 10.1007/s12274-015-0948-y.

[14] F. Igbari, Z. Wang, and L. Liao, “Progress of Lead-Free Halide Double Perovskites,” *Adv Energy Mater*, vol. 9, no. 12, p. 1803150, Mar. 2019, doi: 10.1002/aenm.201803150.

[15] X.-G. Zhao, D. Yang, J.-C. Ren, Y. Sun, Z. Xiao, and L. Zhang, “Rational Design of Halide Double Perovskites for Optoelectronic Applications,” *Joule*, vol. 2, no. 9, pp. 1662–1673, Sep. 2018, doi: 10.1016/j.joule.2018.06.017.

[16] J. C. Dahl *et al.*, “Probing the Stability and Band Gaps of $Cs_2AgInCl_6$ and $Cs_2AgSbCl_6$ Lead-Free Double Perovskite Nanocrystals,” *Chemistry of Materials*, vol. 31, no. 9, pp. 3134–3143, May 2019, doi: 10.1021/acs.chemmater.8b04202.

[17] A. H. Slavney, T. Hu, A. M. Lindenberg, and H. I. Karunadasa, “A Bismuth-Halide Double Perovskite with Long Carrier Recombination Lifetime for Photovoltaic Applications,” *J Am Chem Soc*, vol. 138, no. 7, pp. 2138–2141, Feb. 2016, doi: 10.1021/jacs.5b13294.

[18] F. Igbari, Z. Wang, and L. Liao, “Progress of Lead-Free Halide Double Perovskites,” *Adv Energy Mater*, vol. 9, no. 12, p. 1803150, Mar. 2019, doi: 10.1002/aenm.201803150.

[19] N. Guechi, A. Bouhemadou, S. Bin-Omran, A. Bourzami, and L. Louail, “Elastic, Optoelectronic and Thermoelectric Properties of the Lead-Free Halide Semiconductors Cs2AgBiX 6 (X = Cl, Br): Ab Initio Investigation,” *J Electron Mater*, vol. 47, no. 2, pp. 1533–1545, Feb. 2018, doi: 10.1007/s11664-017-5962-2.

[20] H. Chen, S. Ming, M. Li, B. Wang, and J. Su, "First-Principles Study on the Structure, Electronic and Optical Properties of $Cs_2AgSb_xBi_{1-x}Cl_6$ Double Perovskites," *The Journal of Physical Chemistry C*, vol. 125, no. 20, pp. 11271–11277, May 2021, doi: 10.1021/acs.jpcc.1c03027.

[21] S. Zhao, K. Yamamoto, S. Iikubo, S. Hayase, and T. Ma, "First-principles study of electronic and optical properties of lead-free double perovskites Cs2NaBX6 (B = Sb, Bi; X = Cl, Br, I)," *Journal of Physics and Chemistry of Solids*, vol. 117, pp. 117–121, Jun. 2018, doi: 10.1016/j.jpcs.2018.02.032.

[22] W. Zhou *et al.*, "Lead-Free Small-Bandgap $Cs_2CuSbCl_6$ Double Perovskite Nanocrystals," *J Phys Chem Lett*, vol. 11, no. 15, pp. 6463–6467, Aug. 2020, doi: 10.1021/acs.jpclett.0c01968.

[23] C. Lin, Y. Zhao, Y. Liu, W. Zhang, C. Shao, and Z. Yang, "The bandgap regulation and optical properties of alloyed Cs2NaSbX6 (X=Cl, Br, I) systems with first principle method," *Journal of Materials Research and Technology*, vol. 11, pp. 1645–1653, Mar. 2021, doi: 10.1016/j.jmrt.2021.01.075.

[24] J. Luo *et al.*, "$Cs_2AgInCl_6$ Double Perovskite Single Crystals: Parity Forbidden Transitions and Their Application For Sensitive and Fast UV Photodetectors," *ACS Photonics*, vol. 5, no. 2, pp. 398–405, Feb. 2018, doi: 10.1021/acsphotonics.7b00837.

[25] R. Anbarasan *et al.*, "Exploring the structural, mechanical, electronic, and optical properties of double perovskites of Cs2AgInX6 (X = Cl, Br, I) by first-principles calculations," *J Solid State Chem*, vol. 310, p. 123025, Jun. 2022, doi: 10.1016/j.jssc.2022.123025.

[26] X. Liu *et al.*, "Tunable Dual-Emission in Monodispersed $Sb^{3+}$/$Mn^{2+}$ Codoped $Cs_2NaInCl_6$ Perovskite Nanocrystals through an Energy Transfer Process," *Small*, vol. 16, no. 31, p. 2002547, Aug. 2020, doi: 10.1002/smll.202002547.

[27] A. H. Slavney *et al.*, "Small-Band-Gap Halide Double Perovskites," *Angewandte Chemie International Edition*, vol. 57, no. 39, pp. 12765–12770, Sep. 2018, doi: 10.1002/anie.201807421.

[28] Z. Xiao, K.-Z. Du, W. Meng, J. Wang, D. B. Mitzi, and Y. Yan, "Intrinsic Instability of $Cs_2In(I)M(III)X_6$ (M = Bi, Sb; X = Halogen) Double Perovskites: A Combined Density Functional Theory and Experimental Study," *J Am Chem Soc*, vol. 139, no. 17, pp. 6054–6057, May 2017, doi: 10.1021/jacs.7b02227.

[29] F. Locardi *et al.*, "Colloidal Synthesis of Double Perovskite $Cs_2AgInCl_6$ and Mn-Doped $Cs_2AgInCl_6$ Nanocrystals," *J Am Chem Soc*, vol. 140, no. 40, pp. 12989–12995, Oct. 2018, doi: 10.1021/jacs.8b07983.

[30] L. Zhang *et al.*, "Investigation on lead-free Mn-doped Cs2NaInCl6 double perovskite phosphors and their optical properties," *Opt Mater (Amst)*, vol. 122, p. 111802, Dec. 2021, doi: 10.1016/j.optmat.2021.111802.

[31] M. Zia ur Rehman *et al.*, "First principles study of structural, electronic, elastic and optical properties of Cs2LiTlBr6 and Cs2NaTlBr6," *Mater Sci Semicond Process*, vol. 151, p. 106993, Nov. 2022, doi: 10.1016/j.mssp.2022.106993.

[32] G. Kresse and J. Furthmüller, "Efficient iterative schemes for *ab initio* total-energy calculations using a plane-wave basis set," *Phys Rev B*, vol. 54, no. 16, pp. 11169–11186, Oct. 1996, doi: 10.1103/PhysRevB.54.11169.

[33] G. Kresse and J. Furthmüller, "Efficiency of ab-initio total energy calculations for metals and semiconductors using a plane-wave basis set," *Comput Mater Sci*, vol. 6, no. 1, pp. 15–50, Jul. 1996, doi: 10.1016/0927-0256(96)00008-0.

[34] J. P. Perdew, K. Burke, and M. Ernzerhof, "Generalized Gradient Approximation Made Simple," *Phys Rev Lett*, vol. 77, no. 18, pp. 3865–3868, Oct. 1996, doi: 10.1103/PhysRevLett.77.3865.

[35] J. P. Perdew *et al.*, "Atoms, molecules, solids, and surfaces: Applications of the generalized gradient approximation for exchange and correlation," *Phys Rev B*, vol. 46, no. 11, pp. 6671–6687, Sep. 1992, doi: 10.1103/PhysRevB.46.6671.

[36] P. Hohenberg and W. Kohn, "Inhomogeneous Electron Gas," *Physical Review*, vol. 136, no. 3B, pp. B864–B871, Nov. 1964, doi: 10.1103/PhysRev.136.B864.

[37] P. E. Blöchl, "Projector augmented-wave method," *Phys Rev B*, vol. 50, no. 24, pp. 17953–17979, Dec. 1994, doi: 10.1103/PhysRevB.50.17953.

[38] J. Heyd, G. E. Scuseria, and M. Ernzerhof, "Hybrid functionals based on a screened Coulomb potential," *J Chem Phys*, vol. 118, no. 18, pp. 8207–8215, May 2003, doi: 10.1063/1.1564060.

[39] G. Volonakis *et al.*, "$Cs_2InAgCl_6$: A New Lead-Free Halide Double Perovskite with Direct Band Gap," *J Phys Chem Lett*, vol. 8, no. 4, pp. 772–778, Feb. 2017, doi: 10.1021/acs.jpclett.6b02682.

[40] L. R. Morss, M. Siegal, L. Stenger, and N. Edelstein, "Preparation of cubic chloro complex compounds of trivalent metals: Cs2NaMCl6," *Inorg Chem*, vol. 9, no. 7, pp. 1771–1775, Jul. 1970, doi: 10.1021/ic50089a034.

[41] I. Khan, Shahab, I. U. Haq, A. Ali, Z. Ali, and I. Ahmad, "Elastic and Optoelectronic Properties of Cs2NaMCl6 (M = In, Tl, Sb, Bi)," *J Electron Mater*, vol. 50, no. 2, pp. 456–466, Feb. 2021, doi: 10.1007/s11664-020-08603-y.

[42] F. Birch, "Finite Elastic Strain of Cubic Crystals," *Physical Review*, vol. 71, no. 11, pp. 809–824, Jun. 1947, doi: 10.1103/PhysRev.71.809.

[43] V. M. Goldschmidt, "Die Gesetze der Krystallochemie," *Naturwissenschaften*, vol. 14, no. 21, pp. 477–485, May 1926, doi: 10.1007/BF01507527.

[44] S. Schneider and R. Hoppe, "�ber neue Verbindungen Cs2NaMF6 und K2NaMF6 sowie �ber Cs2KMnF6," *Zeitschrift f�r anorganische und allgemeine Chemie*, vol. 376, no. 3, pp. 268–276, Sep. 1970, doi: 10.1002/zaac.19703760309.

[45] L. R. Morss, M. Siegal, L. Stenger, and N. Edelstein, "Preparation of cubic chloro complex compounds of trivalent metals: Cs2NaMCl6," *Inorg Chem*, vol. 9, no. 7, pp. 1771–1775, Jul. 1970, doi: 10.1021/ic50089a034.

[46] M. Zia ur Rehman *et al.*, "First principles study of structural, electronic, elastic and optical properties of Cs2LiTlBr6 and Cs2NaTlBr6.," *Mater Sci Semicond Process*, vol. 151, p. 106993, Nov. 2022, doi: 10.1016/j.mssp.2022.106993.

[47] S. Nair, M. Deshpande, V. Shah, S. Ghaisas, and S. Jadkar, "$Cs_2TlBiI_6$: a new lead-free halide double perovskite with direct band gap," *Journal of Physics: Condensed Matter*, vol. 31, no. 44, p. 445902, Nov. 2019, doi: 10.1088/1361-648X/ab32a5.

[48] C. J. Bartel *et al.*, "Physical descriptor for the Gibbs energy of inorganic crystalline solids and temperature-dependent materials chemistry," *Nat Commun*, vol. 9, no. 1, p. 4168, Oct. 2018, doi: 10.1038/s41467-018-06682-4.

[49] E. Tenuta, C. Zheng, and O. Rubel, "Thermodynamic origin of instability in hybrid halide perovskites," *Sci Rep*, vol. 6, no. 1, p. 37654, Nov. 2016, doi: 10.1038/srep37654.

[50] N. Marzari, A. Ferretti, and C. Wolverton, "Electronic-structure methods for materials design," *Nat Mater*, vol. 20, no. 6, pp. 736–749, Jun. 2021, doi: 10.1038/s41563-021-01013-3.

[51] G. Volonakis *et al.*, "$Cs_2InAgCl_6$: A New Lead-Free Halide Double Perovskite with Direct Band Gap," *J Phys Chem Lett*, vol. 8, no. 4, pp. 772–778, Feb. 2017, doi: 10.1021/acs.jpclett.6b02682.

[52] K. Wang, Y. He, M. Zhang, J. Shi, and W. Cai, "Promising Lead-Free Double-Perovskite Photovoltaic Materials $Cs_2MM'Br_6$ (M = Cu, Ag, and Au; M′ = Ga, In, Sb, and Bi) with an

Ideal Band Gap and High Power Conversion Efficiency," *The Journal of Physical Chemistry C*, vol. 125, no. 38, pp. 21160–21168, Sep. 2021, doi: 10.1021/acs.jpcc.1c05699.

[53] N. A. Noor *et al.*, "Analysis of direct band gap A2ScInI6 (A=Rb, Cs) double perovskite halides using DFT approach for renewable energy devices," *Journal of Materials Research and Technology*, vol. 13, pp. 2491–2500, Jul. 2021, doi: 10.1016/j.jmrt.2021.05.080.

[54] S. Nair, M. Deshpande, V. Shah, S. Ghaisas, and S. Jadkar, "$Cs_2TlBiI_6$ : a new lead-free halide double perovskite with direct band gap," *Journal of Physics: Condensed Matter*, vol. 31, no. 44, p. 445902, Nov. 2019, doi: 10.1088/1361-648X/ab32a5.

[55] Md. N. Islam, J. Podder, T. Saha, and P. Rani, "Semiconductor to metallic transition under induced pressure in $Cs_2AgBiBr_6$ double halide perovskite: a theoretical DFT study for photovoltaic and optoelectronic applications," *RSC Adv*, vol. 11, no. 39, pp. 24001–24012, 2021, doi: 10.1039/D1RA03161A.

[56] D. Behera and S. K. Mukherjee, "Optoelectronics and Transport Phenomena in Rb2InBiX6 (X = Cl, Br) Compounds for Renewable Energy Applications: A DFT Insight," *Chemistry (Easton)*, vol. 4, no. 3, pp. 1044–1059, Sep. 2022, doi: 10.3390/chemistry4030071.

[57] A. Mattoni and C. Caddeo, "Dielectric function of hybrid perovskites at finite temperature investigated by classical molecular dynamics," *J Chem Phys*, vol. 152, no. 10, p. 104705, Mar. 2020, doi: 10.1063/1.5133064.

[58] T. Ghrib *et al.*, "A new lead free double perovskites K2Ti(Cl/Br)6; a promising materials for optoelectronic and transport properties; probed by DFT," *Mater Chem Phys*, vol. 264, p. 124435, May 2021, doi: 10.1016/j.matchemphys.2021.124435.

[59] L.-K. Gao and Y.-L. Tang, "Theoretical Study on the Carrier Mobility and Optical Properties of $CsPbI_3$ by DFT," *ACS Omega*, vol. 6, no. 17, pp. 11545–11555, May 2021, doi: 10.1021/acsomega.1c00734.

[60] N. Hasan, M. Arifuzzaman, and A. Kabir, "Structural, elastic and optoelectronic properties of inorganic cubic $FrBX_3$ (B = Ge, Sn; X = Cl, Br, I) perovskite: the density functional theory approach," *RSC Adv*, vol. 12, no. 13, pp. 7961–7972, 2022, doi: 10.1039/D2RA00546H.

[61] I. B. Ogunniranye, T. Atsue, and O. E. Oyewande, "Structural and optoelectronic behavior of the copper-doped <math> <msub> <mrow> <mi>Cs</mi> </mrow> <mn>2</mn> </msub> <msub> <mrow> <mi>AgInCl</mi> </mrow> <mn>6</mn> </msub> </math> double perovskite: A density functional theory investigation," *Phys Rev B*, vol. 103, no. 2, p. 024102, Jan. 2021, doi: 10.1103/PhysRevB.103.024102.

[62] H. Yu, Y. Peng, Y. Yang, and Z.-Y. Li, "Plasmon-enhanced light–matter interactions and applications," *NPJ Comput Mater*, vol. 5, no. 1, p. 45, Dec. 2019, doi: 10.1038/s41524-019-0184-1.

[63] K. Roccapriore, A. Bozhko, G. Nazarikov, V. Drachev, and A. Krokhin, "Surface plasmon at a metal-dielectric interface with an epsilon-near-zero transition layer," *Phys Rev B*, vol. 103, no. 16, p. L161404, Apr. 2021, doi: 10.1103/PhysRevB.103.L161404.

[64] L. Guo, G. Tang, and J. Hong, "Mechanical Properties of Formamidinium Halide Perovskites $FABX_3$ ($FA=CH(NH_2)_2$; B=Pb, Sn; X=Br, I) by First-Principles Calculations *," *Chinese Physics Letters*, vol. 36, no. 5, p. 056201, May 2019, doi: 10.1088/0256-307X/36/5/056201.

[65] M. Hussain, M. Rashid, A. Ali, M. F. Bhopal, and A. S. Bhatti, "Systematic study of optoelectronic and transport properties of cesium lead halide (Cs2PbX6; X=Cl, Br, I) double perovskites for solar cell applications," *Ceram Int*, vol. 46, no. 13, pp. 21378–21387, Sep. 2020, doi: 10.1016/j.ceramint.2020.05.235.

[66] S. Sun *et al.*, "Factors Influencing the Mechanical Properties of Formamidinium Lead Halides and Related Hybrid Perovskites," *ChemSusChem*, vol. 10, no. 19, pp. 3740–3745, Oct. 2017, doi: 10.1002/cssc.201700991.

[67] T. Skośkiewicz, “Thermal Conductivity at Low Temperatures,” in *Encyclopedia of Condensed Matter Physics*, Elsevier, 2005, pp. 159–164. doi: 10.1016/B0-12-369401-9/01168-2.
[68] V. D. Natsik and P. P. Pal-val, “Low-temperature dislocation internal friction in crystals,” in *Fundamental Aspects of Dislocation Interactions*, Elsevier, 1993, pp. 312–315. doi: 10.1016/B978-1-4832-2815-0.50051-6.
[69] C. Li and Z. Wang, “Computational modelling and ab initio calculations in MAX phases – I,” in *Advances in Science and Technology of Mn+1AXn Phases*, Elsevier, 2012, pp. 197–222. doi: 10.1533/9780857096012.197.